\documentclass[conference]{IEEEtran}
\IEEEoverridecommandlockouts

\usepackage{cite}
\usepackage{amsmath,amssymb,amsfonts}
\usepackage{algorithmic}
\usepackage{graphicx}
\usepackage{textcomp}
\usepackage{enumitem}
\usepackage{xcolor}
\usepackage{booktabs}
\usepackage{stfloats}
\usepackage{multirow}
\usepackage{makecell}
\usepackage{colortbl}
\usepackage{subcaption}

\definecolor{rankbad}{HTML}{E03131}    
\definecolor{rankmid}{HTML}{F08C00}    
\definecolor{rankgood}{HTML}{2F9E44}   
\definecolor{rankgreat}{HTML}{1971C2}  
\definecolor{rowshade}{HTML}{F5F5F5}   
\def\BibTeX{{\rm B\kern-.05em{\sc i\kern-.025em b}\kern-.08em
    T\kern-.1667em\lower.7ex\hbox{E}\kern-.125emX}}
\begin{document}
\bstctlcite{IEEE:BSTcontrol}
\title{Configuration Over Selection: Hyperparameter Sensitivity Exceeds Model Differences in Open-Source LLMs for RTL Generation
}
\author{%
Minghao~Shao$^\dag$$^\ddag$
Zeng~Wang$^\dag$,
Weimin~Fu$^\P$,
Xiaolong~Guo$^\P$,
Johann~Knechtel$^\ddag$,\\
Ozgur~Sinanoglu$^\ddag$,
Ramesh~Karri$^\dag$,
Muhammad~Shafique$^\ddag$
\\
\IEEEauthorblockA{
$^\dag$NYU Tandon School of Engineering, USA \ 
$^\ddag$NYU Abu Dhabi, UAE \ 
$^\P$Kansas State University, USA\\
\normalsize{Email: \{shao.minghao, zw3464, johann, ozgursin, rkarri, muhammad.shafique\}@nyu.edu}\\
\normalsize{\{weiminf, guoxiaolong\}}@ksu.edu}
}

\maketitle

\begin{abstract}
Benchmarking of open-source LLMs for hardware design focuses on \emph{which} LLMs to use, while treating inference-time decoding configuration as a secondary concern. This work shows that it matters more \emph{how} an LLM is configured than which model is selected. Benchmarking 26 open-source LLMs on VerilogEval and RTLLM with synthesis-in-the-loop evaluation, the study first maps the current capability landscape and then conducts an extensive 108-configuration hyperparameter sweep on three prominent models. The sweep reveals absolute pass-rate gaps of up to 25.5\% between the best and worst settings for the same LLM, which is 5x larger than the average spread observed across various model families under their respective default configurations. Ranking all configurations by Spearman's $\rho$ across the two benchmark suites yields near-zero correlation, demonstrating that optimal configurations do not transfer. These results show that benchmarking conducted under default hyperparameters confounds model capabilities with configuration effects. Realizing the full potential of open-source LLMs for RTL generation requires architecture and benchmark aware hyperparameter selection, as enabled by the proposed methodology.
\end{abstract}

\section{Introduction}

Large language models (LLMs) have shown significant capability in automated hardware description generation~\cite{thakur2022benchmarkinglargelanguagemodels, wang2024llms}. Register-transfer level (RTL) design remains a bottleneck in semiconductor manufacturing, requiring engineers to balance functional accuracy with timing and area constraints. Although AI approaches can accelerate chip development, passing software simulations alone is insufficient: for machine-generated hardware to replace golden modules, it must survive the physical synthesis pipeline, making synthesis-backed validation a rigorous measure of design quality~\cite{abdelatty2025pluto}.

Recent evaluations of LLM-aided RTL generation indicate that commercial LLMs maintain higher accuracy than their open-source counterparts~\cite{cui2024origen}, yet open-source LLMs offer distinct practical advantages. Open-weights LLMs enable cost-free local deployment without recurring API costs and allow domain-specific fine-tuning that is unavailable with commercial equivalents~\cite{liu2024rtlcoder}. For the semiconductor industry, where intellectual property protection is a concern~\cite{wang2025verileaky}, open-source LLMs enable isolated on-premise inference, removing the risk of sensitive design data reaching third-party servers.

\begin{figure}[htbp]
    \centering
    \includegraphics[width=\columnwidth]{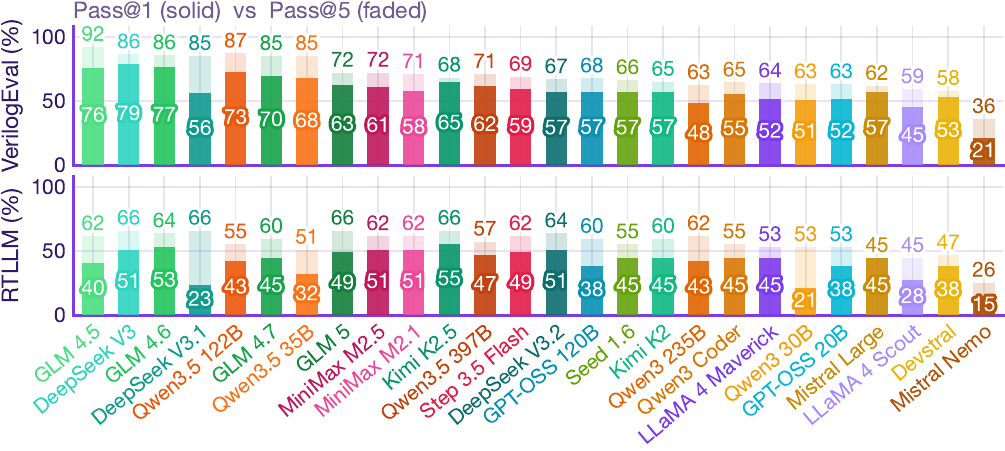}
    \caption{Pass rates of 26 open-source LLMs on VerilogEval and RTLLM.}
    \label{fig:benchmark_pass_rates}
\end{figure}

Despite these advantages, the performance ceilings and optimization strategies for open-source models remain underexplored~\cite{lu2024openllmrtl}. A critical but largely overlooked factor is inference-time decoding configuration. Existing evaluations compare models under default or unspecified hyperparameters, implicitly treating configuration as a minor implementation detail. Studies on general code generation show that decoding choices such as temperature and top-p can materially change functional success rates~\cite{zhu2024hot}. This raises a key question: when model~A outperforms model~B on an RTL benchmark, how much of that gap reflects genuine model capability, and how much is an artifact of the hyperparameters used?

To answer this question, we evaluate 26 open-source models across VerilogEval and RTLLM with synthesis-in-the-loop assessment, first establishing a performance landscape to quantify inter-model differences, and then conducting a 108-configuration hyperparameter sweep on three architectures. The sweep shows that the same model's pass rate varies by up to 25.5\% depending solely on decoding settings, a configuration-induced variation that exceeds the pass-rate spread between entire model families observed in the landscape evaluation. Spearman's rank correlation of configuration rankings across the two benchmarks is near zero for all three models, indicating that a configuration optimized on one benchmark has no predictive value for performance on another.

\begin{figure*}[t]
  \centering
  \includegraphics[width=1.04\linewidth]{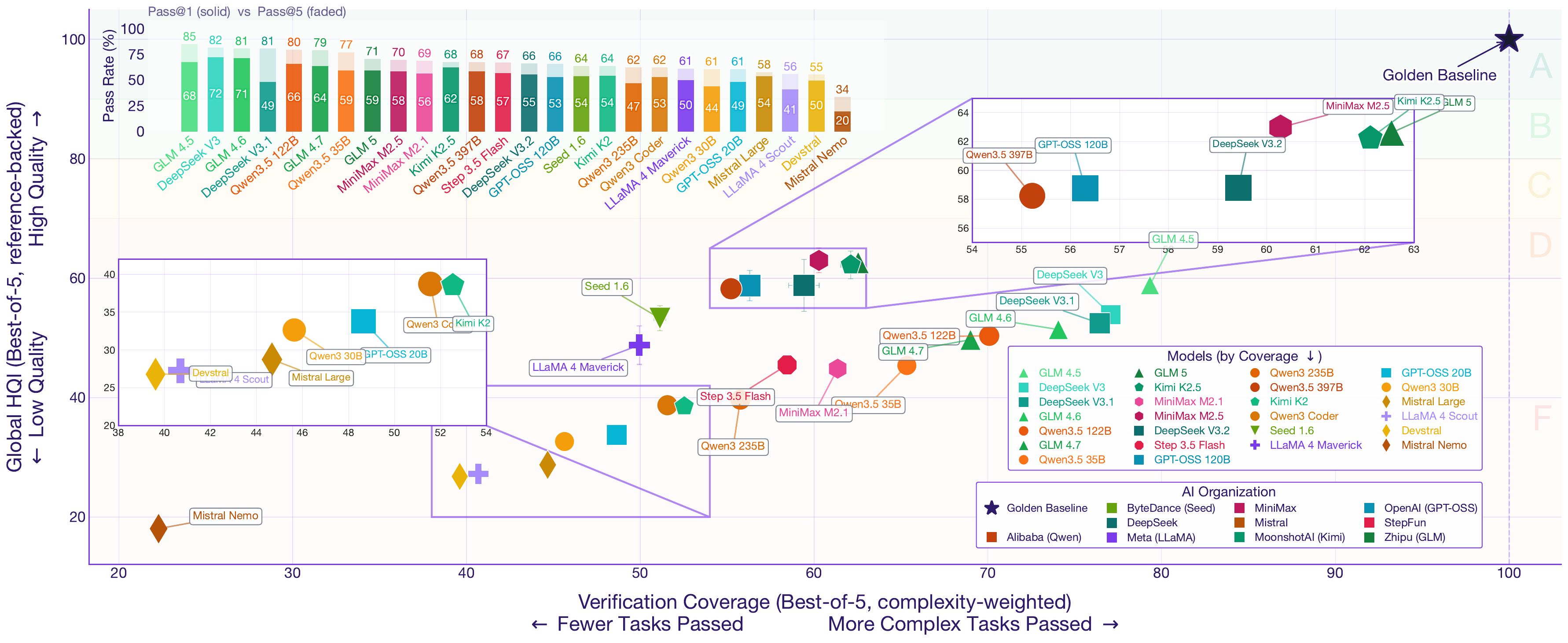}
    \caption{Pass-rate and HQI landscape for 26 language models. Top: pass@1 (solid) and pass@5 (faded) sorted by descending pass@5. Bottom: Complexity-weighted verification coverage versus Global HQI under synthesis-in-the-loop evaluation.}
  \label{fig:coverage_gpa}
\end{figure*}

In summary, our primary contributions are:
\begin{itemize}[leftmargin=*]
    \item We benchmark 26 open-source LLMs with synthesis-in-the-loop evaluation to establish the magnitude of inter-model performance differences, providing a baseline against which the impact of hyperparameters can be measured.
    \item We demonstrate through a 108-configuration sweep that hyperparameter-induced performance variation (up to 25.5\%) can exceed inter-model variation, establishing inference-time configuration as a primary determinant of open-source LLM performance for RTL generation.
    \item We reveal that optimal hyperparameter settings are architecture-specific and benchmark-specific (Spearman's $\rho \approx 0$ across benchmarks), ruling out universal ``recommended configurations'' and motivating task-aware tuning.
\end{itemize}

\section{Background}

Large language models have attracted increasing interest in EDA, spanning hardware generation, design optimization, hardware security, and logic synthesis~\cite{wang2025netdetox}. Their success in natural-language-to-code tasks~\cite{shao2024survey}, reinforced by execution-based benchmarks such as HumanEval and MBPP~\cite{chen2021evaluating, austin2021program}, encouraged researchers to target synthesizable RTL~\cite{thakur2022benchmarkinglargelanguagemodels}. RTL generation is harder to evaluate than software synthesis: a candidate design must be syntactically valid, functionally correct under simulation, and useful for downstream implementation~\cite{lu2024rtllm}. VerilogEval introduced automated simulation-based checking with pass@k as a central metric~\cite{liu2023verilogeval}; RTLLM extended the scope to natural-language-to-RTL generation with syntax, functionality, and design-quality assessment~\cite{lu2024rtllm}; and OpenLLM-RTL expanded the benchmarks with updated tasks and larger training corpora~\cite{lu2024openllmrtl}.

Model-focused efforts including VeriGen and RTLCoder showed that domain specialization can substantially strengthen HDL generation, with even relatively compact models achieving competitive benchmark results~\cite{thakur2024verigen, liu2024rtlcoder}. Agentic approaches now extend RTL generation beyond one-shot prompting by incorporating iterative refinement, tool use, and multi-step reasoning into the design loop~\cite{zhao2025mage}. Current evaluations still follow early code benchmark logic, ending at simulation pass rates. This leaves a significant gap, as RTL is only an intermediate representation whose value ultimately depends on post-synthesis implementation quality~\cite{garcia2025turtle}. This limitation has motivated re-evaluations of hardware code benchmarks, which emphasize how strongly reported performance depends on benchmark design and evaluation protocol~\cite{pinckney2025revisiting}. Open-source EDA infrastructure now enables richer assessment: Yosys provides a widely used framework for RTL synthesis, and the Nangate45 open cell library offers a reproducible technology context for academic quality-of-results studies~\cite{wolf2013yosys, nangate45}, enabling the synthesis-in-the-loop evaluation.

Open-source models therefore deserve attention, not simply as cheaper stand-ins for proprietary systems. Recent work suggests that their performance ceiling can change substantially with model family, training data, and domain specialization~\cite{liu2024rtlcoder, cui2024origen}. Studies on code generation show that inference-time decoding choices such as temperature and top-p sampling can materially change functional success rates~\cite{zhu2024hot}, but existing RTL evaluations typically report results under default or unspecified hyperparameters, making it impossible to determine whether observed model rankings reflect intrinsic capability differences or incidental configuration choices. For RTL generation, this means a fair comparison of open-weight models must distinguish the capability of the pretrained model from gains or losses introduced by runtime configuration.

\section{Experiment Setup}

\paragraph{Model Selection}
To ensure broad model coverage, we evaluate 26 open-source models spanning eight families: Qwen (Qwen-3.5 397B/122B/35B, Qwen-3 235B/30B/Coder), GLM (GLM 5/4.7/4.6/4.5), DeepSeek (V3.2/V3.1/V3), Mistral (Large/Nemo/Devstral), Kimi (K2.5/K2), MiniMax (M2.5/M2.1), LLaMA (4 Maverick/4 Scout), and GPT-OSS (120B/20B), as well as Seed~1.6 and Step~3.5~Flash.

\paragraph{Benchmarks}
Two established RTL generation benchmarks, VerilogEval~\cite{liu2023verilogeval} and RTLLM~\cite{lu2024rtllm}, are used, which provide complementary evaluation settings and reduce the risk that conclusions are driven by a single task style.

\paragraph{Hyperparameter Setup}
\begin{figure*}[htbp]
  \centering
  \includegraphics[width=1.02\linewidth]{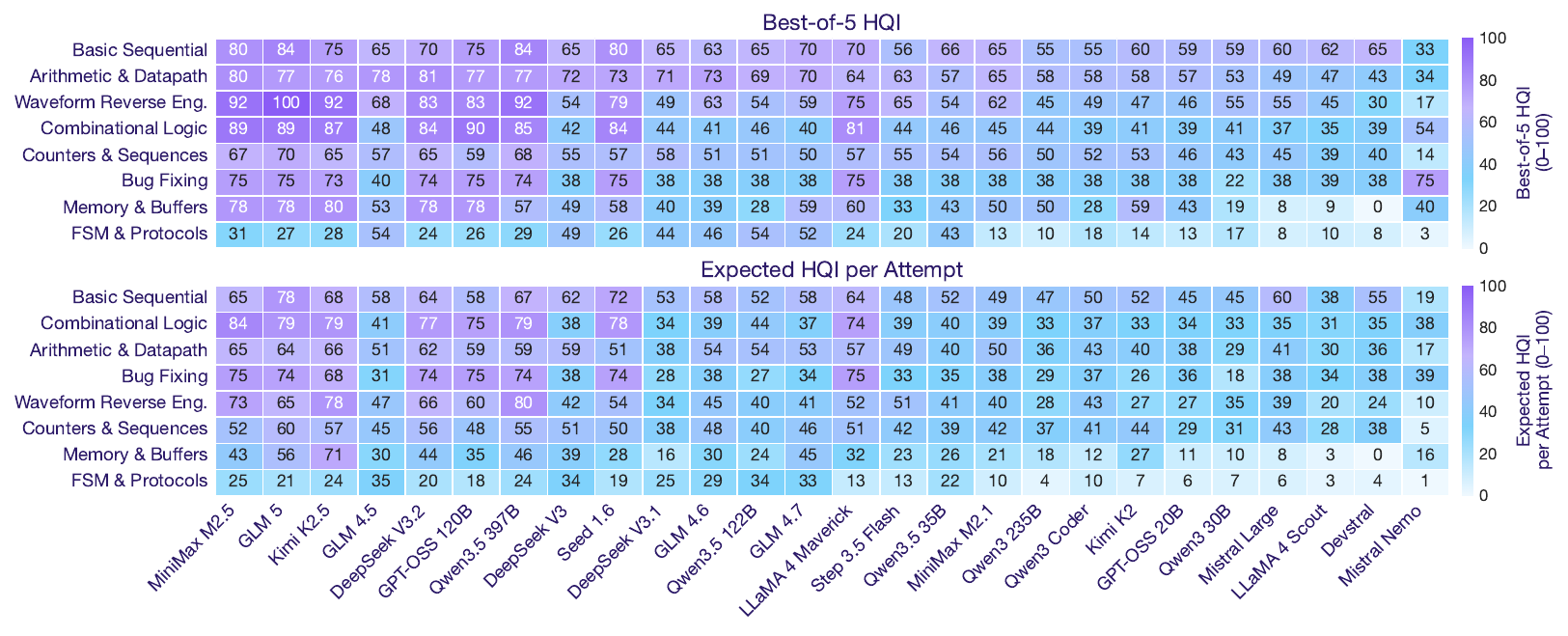}
  \caption{HQI across eight hardware categories and 26 models. Models are ordered left-to-right by Global HQI; categories top-to-bottom by top models' average scores. Top: best-of-five capability ceiling. Bottom: single-attempt deployment quality.}
  \label{fig:heatmap_combined}
\end{figure*}

Three representative models are selected for the hyperparameter sweep: GPT-OSS~120B, Qwen-3.5~397B, and GLM-5. These are high-performing members of distinct families whose pilot experiments (default vs.\ modified temperature) revealed different levels of sensitivity, enabling a test of whether hyperparameter effects are architecture-dependent. We vary four decoding parameters: \texttt{temperature} (distribution entropy), \texttt{top\_p} (probability-mass truncation), \texttt{repetition\_penalty} (token reuse suppression), and \texttt{presence\_penalty} (novelty bias). Together they cover the principal dimensions of decoding behavior: determinism, search-space restriction, repetition, and diversity.

\paragraph{Metrics}
We report pass@1 through pass@5 on both benchmarks as primary quality metrics, alongside HQI, expected HQI, and complexity-weighted coverage for post-synthesis design quality. The Hardware Quality Index (HQI) follows~\cite{fu2026synthesisintheloopevaluationllmsrtl}: a design that fails any evaluation gate receives $\mathrm{HQI}=0$. For a passing design on task $t$, let $\hat{a}$, $\hat{d}$, $\hat{w}$ denote its post-synthesis area, delay, and warning count, and $a^{*}_{t}$, $d^{*}_{t}$, $w^{*}_{t}$ the golden-reference values. The normalized cost is:
\begin{equation}
\textstyle
\mathrm{cost} = 0.5\,\frac{\hat{a}}{a^{*}_{t}} + 0.5\,\frac{\hat{d}}{d^{*}_{t}} + 0.1\,\max\!\bigl(0,\;\hat{w}-w^{*}_{t}\bigr),
\label{eq:hqi_cost}
\end{equation}
and the attempt is scored as $\mathrm{HQI}=\min(100/\mathrm{cost},\;100)$, where 100 indicates parity with the golden reference. Per-task scores are aggregated using complexity weights $C_t$ derived from AST dependency-edge counts: Global HQI takes the best-of-five ceiling per task, while Expected HQI averages over all five attempts. We also report deployment efficiency metrics (cost per task, throughput, time-to-first-token, and completion tokens). For the hyperparameter study, we report pass@1 and pass@5 on both benchmarks; the gap between the best and worst settings directly quantifies how much configuration choice affects performance relative to model choice.


\paragraph{Framework Implementation}
The evaluation is a unified Python pipeline that merges VerilogEval and RTLLM, removes duplicated tasks by  RTL matching, queries each prompt through an OpenAI-compatible API, and extract the first fenced Verilog block as the candidate design. Generated RTL is evaluated with Icarus~Verilog for syntax checking and simulation and Yosys with ABC for synthesis and structural analysis. The pipeline records syntax validity, synthesizability, simulation pass, and synthesis-derived statistics for each task.

\section{Evaluation Results}

\subsection{Performance Landscape}

\subsubsection{Pass Rate vs. HQI}
Figure~\ref{fig:coverage_gpa} ranks all 26 models by pass@5 alongside pass@1, showing a consistent gap between the two metrics across all evaluated models. Even models with high synthesis quality such as GLM~5 exhibit a notable decline from pass@5 to pass@1, indicating that single-attempt deployment remains unreliable across the model pool. At the model-family level, DeepSeek, GLM, and Qwen dominate both pass-rate rankings and Global~HQI, while the Mistral family clusters toward the lower end of both axes. The bottom panel shows that pass rate and synthesis quality do not always align: DeepSeek~V3.2 and GPT-OSS~120B achieve higher Global~HQI relative to their coverage, indicating strong post-synthesis quality on the tasks they do solve. Generational progression does not uniformly improve RTL capability either; within the GLM family, newer releases trade coverage breadth for design depth, a distinction that pass-rate metrics alone do not capture. Under default configurations, the pass@5 difference between the top-ranked and mid-ranked model families (e.g., DeepSeek~V3.2 vs.\ LLaMA~4~Maverick) is approximately 15--20 percentage points. Section~IV-B shows that a single model's performance can swing by up to 25.5\% under different hyperparameter settings alone, placing configuration-induced variation on the same scale as inter-family differences.



\subsubsection{Category-Level Analysis}

Figure~\ref{fig:heatmap_combined} decomposes synthesis quality across eight hardware categories. Basic Sequential, Combinational Logic, and Waveform Reverse Engineering score highest, with top models reaching near-golden levels, whereas FSM \& Protocols and Memory \& Buffers remain difficult even for the strongest models. Both panels reveal a systematic reliability drop from best-of-five to per-attempt evaluation, widening most in these already-difficult categories. This compounding effect suggests the hardest categories are most exposed to hyperparameter misconfiguration, though confirming requires per-category analysis not conducted here.

\begin{figure*}[t]
    \centering
    \begin{subfigure}{0.49\textwidth}
        \centering
        \includegraphics[width=\linewidth]{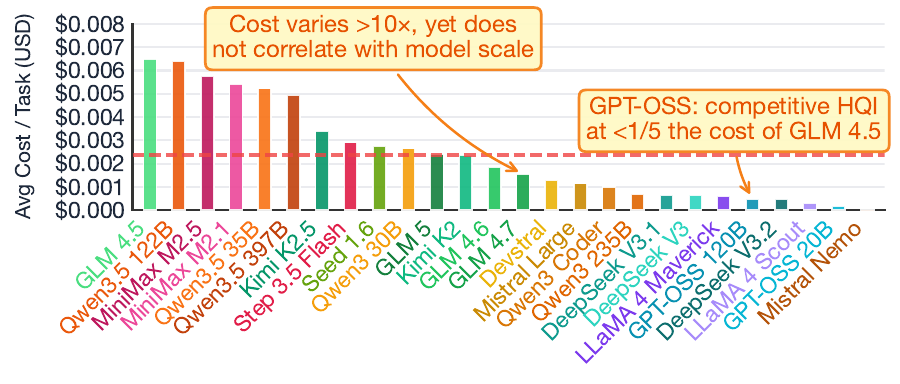}
        \caption{Cost}
        \label{fig:cost}
    \end{subfigure}\hfill
    \begin{subfigure}{0.49\textwidth}
        \centering
        \includegraphics[width=\linewidth]{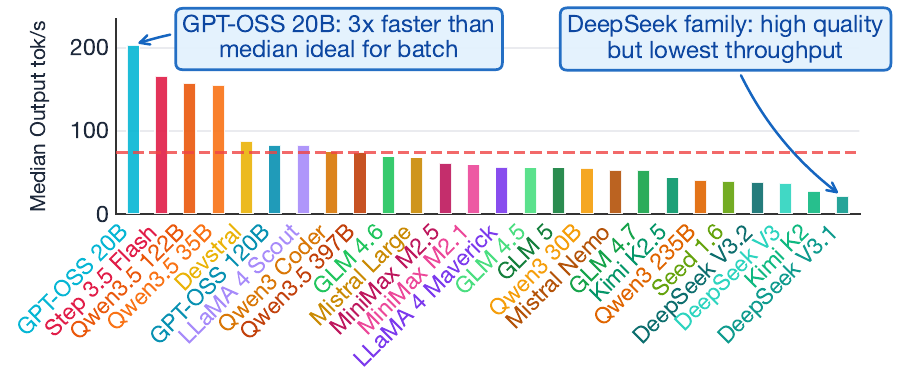}
        \caption{Throughput}
        \label{fig:throughput}
    \end{subfigure}
    
    \medskip
    
    \begin{subfigure}{0.49\textwidth}
        \centering
        \includegraphics[width=\linewidth]{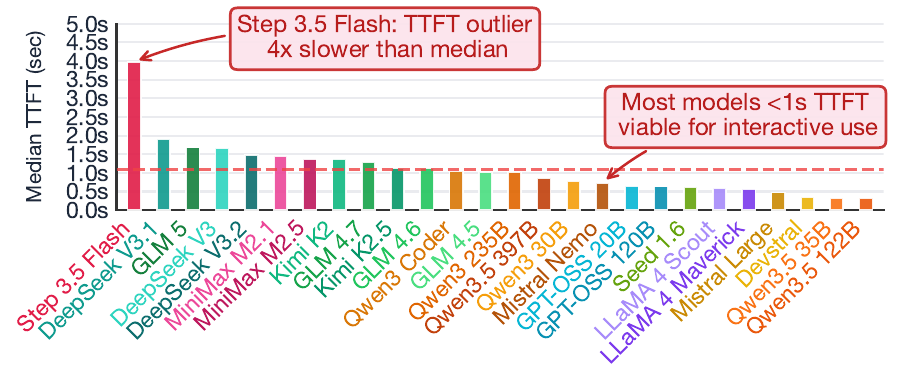}
        \caption{TTFT}
        \label{fig:ttft}
    \end{subfigure}\hfill
    \begin{subfigure}{0.49\textwidth}
        \centering
        \includegraphics[width=\linewidth]{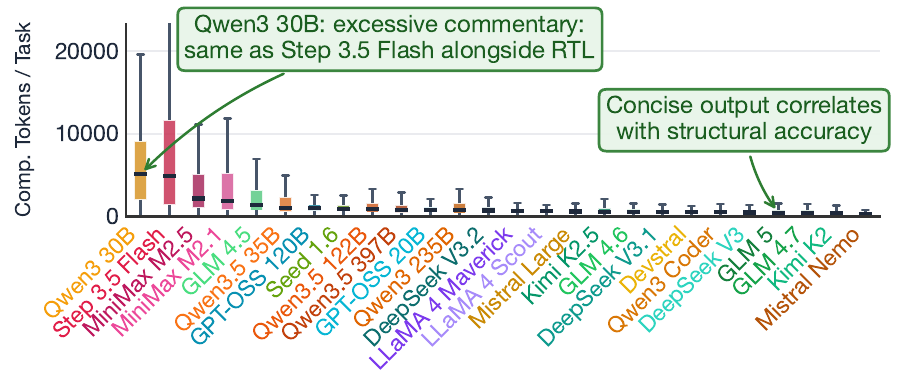}
        \caption{Verbosity}
        \label{fig:verbosity}
    \end{subfigure}
    
    \caption{Inference performance metrics.}
    \label{fig:inference_metrics}
\end{figure*}

\subsubsection{Efficiency Evaluation}

Figure~\ref{fig:inference_metrics} profiles four operational metrics. Cost per task (Figure~\ref{fig:cost}) varies by over an order of magnitude and does not correlate with model scale: GPT-OSS family achieves competitive HQI at substantially lower cost than GLM~4.5 or Qwen-3.5~397B.



Throughput and latency (Figures~\ref{fig:throughput} and~\ref{fig:ttft}) show that interactive and batch workloads may favor different architectures, but these metrics are primarily determined by the serving infrastructure and are orthogonal to the hyperparameter effects studied in Section~IV-B.
Verbosity (Figure~\ref{fig:verbosity}) is more directly related to decoding configuration: models like Qwen3~30B and Step~3.5~Flash produce high token counts with large variance, indicating excessive natural-language commentary alongside the generated RTL. GLM~5 and Kimi~K2.5 maintain high synthesis quality with minimal verbosity, confirming that concise output correlates with structural accuracy. Because decoding hyperparameters such as temperature and penalty terms influence output length, verbosity may interact with the configuration effects examined in Section~IV-B, though the present study does not isolate this relationship.

\subsection{Hyperparameter Sweeping}

\newcommand{\scorehp}[2]{%
  #1\,\scriptsize\textcolor{gray!80}{(#2)}%
}

\begin{table}[t]
\centering
\caption{Performance gap ($\Delta$) between the best and worst hyperparameter settings for three open-source models on VerilogEval and RTLLM. Each setting is a tuple of (temperature, top\_p, repetition\_penalty, presence\_penalty).}
\label{tab:hp_gap_detailed}

\setlength{\tabcolsep}{2pt}       
\renewcommand{\arraystretch}{1.0} 
\footnotesize

\resizebox{\columnwidth}{!}{
\begin{tabular}{@{} l l c cc >{\bfseries}c @{}}
\toprule
\textbf{Model} & \textbf{Benchmark} & \textbf{Metric} & \textbf{Best Score (HP)} & \textbf{Worst Score (HP)} & \textbf{Gap ($\Delta$)} \\
\midrule

\multirow{4}{*}{\begin{tabular}[c]{@{}l@{}}\textbf{GPT-OSS}\\ \scriptsize{(120B)}\end{tabular}}
 & \multirow{2}{*}{VerilogEval}
 & pass@1 & \scorehp{0.626}{0.4,1,1,-1} & \scorehp{0.503}{1.2,1,1.1,0} & 0.123 \\
 & & pass@5 & \scorehp{0.710}{0.4,1,1.1,-1} & \scorehp{0.658}{1.2,1,1.2,1} & 0.052 \\
\cmidrule(l){2-6}
 & \multirow{2}{*}{RTLLM}
 & pass@1 & \scorehp{0.575}{0,0.7,1,-1} & \scorehp{0.319}{0.8,1,1,0} & 0.255 \\
 & & pass@5 & \scorehp{0.681}{0.8,1,1.1,-1} & \scorehp{0.553}{0.4,0.4,1,0} & 0.128 \\
\midrule

\multirow{4}{*}{\begin{tabular}[c]{@{}l@{}}\textbf{Qwen-3.5}\\ \scriptsize{(397B)}\end{tabular}}
 & \multirow{2}{*}{VerilogEval}
 & pass@1 & \scorehp{0.845}{0,0.4,1.1,-1} & \scorehp{0.600}{1.2,1,1.2,0} & 0.245 \\
 & & pass@5 & \scorehp{0.942}{1.2,0.7,1.1,1} & \scorehp{0.858}{1.2,1,1.2,1} & 0.084 \\
\cmidrule(l){2-6}
 & \multirow{2}{*}{RTLLM}
 & pass@1 & \scorehp{0.596}{1.2,0.4,1,0} & \scorehp{0.362}{0.4,1,1.1,0} & 0.234 \\
 & & pass@5 & \scorehp{0.681}{0,0.7,1.2,1} & \scorehp{0.575}{0.8,1,1.2,-1} & 0.106 \\
\midrule

\multirow{4}{*}{\textbf{GLM-5}}
 & \multirow{2}{*}{VerilogEval}
 & pass@1 & \scorehp{0.671}{0,0.4,1.1,0} & \scorehp{0.574}{1.2,1,1.2,0} & 0.097 \\
 & & pass@5 & \scorehp{0.736}{0.8,0.7,1.2,0} & \scorehp{0.684}{0.8,0.4,1.1,1} & 0.052 \\
\cmidrule(l){2-6}
 & \multirow{2}{*}{RTLLM}
 & pass@1 & \scorehp{0.617}{0.8,0.7,1,0} & \scorehp{0.447}{1.2,1,1,0} & 0.170 \\
 & & pass@5 & \scorehp{0.702}{1.2,1,1.2,1} & \scorehp{0.596}{0,0.7,1.2,1} & 0.106 \\
\bottomrule
\end{tabular}%
}
\end{table}

\subsubsection{Performance Gap Under Hyperparameter Sweeping}

Table~\ref{tab:hp_gap_detailed} quantifies the central finding: across the three models, the absolute quality gap between the best and worst hyperparameter settings reaches 25.5\%, matching or exceeding the 15--20 percentage-point spread between top-tier and mid-tier model families observed in Section~IV-A. A well-tuned GPT-OSS~120B on RTLLM (pass@1 = 0.575) outperforms a poorly configured Qwen-3.5~397B (pass@1 = 0.362) despite having less than one-third the parameters. The sensitivity concentrates on first-attempt correctness: pass@1 gaps exceed pass@5 counterparts for all three models, indicating that suboptimal configurations impair immediate solution accuracy rather than erasing underlying hardware knowledge.


Model sensitivity varies: Qwen-3.5~397B and GPT-OSS~120B show double-digit fluctuations across most benchmark metrics, while GLM-5 maintains narrower gaps. This implies a robustness--ceiling trade-off: GLM-5 offers safer defaults but a lower peak, whereas the others reach higher ceilings at the cost of greater configuration dependence.

\subsubsection{Default Configuration Positioning}
Table~\ref{tab:default-position} reveals that default configurations are frequently far from optimal, in the worst cases occupying the very bottom of the 109-setting ranking (108 sweeping + default). Qwen-3.5 397B is the clearest case: its default ranks last on VerilogEval pass@5 and near-last on pass@1, leaving over 22\% of achievable performance unrealized. GPT-OSS 120B tells a similar story on RTLLM, where its default ranks 96th on pass@1 with nearly 19\% left on the table. GLM-5 is the most robust, reaching the top five on VerilogEval pass@5, though it still drops to rank 82 on RTLLM pass@1. These results concretely indict current evaluation practice: the majority of default configurations are not merely suboptimal but actively detrimental for RTL generation, in multiple documented cases placing models at their empirical floor and obscuring the intrinsic capability that these models otherwise possess.

\newcommand{\pctcell}[2]{%
  \ifnum#2>89
    \textcolor{rankgreat}{\textbf{#1}}%
  \else\ifnum#2>60
    \textcolor{rankgood}{#1}%
  \else\ifnum#2>25
    \textcolor{rankmid}{#1}%
  \else
    \textcolor{rankbad}{\textbf{#1}}%
  \fi\fi\fi
}
\begin{table}[t]
\centering
\caption{Default (1.0, 1.0, 1.0, 0) configuration rank among 109 hyperparameter settings. \textit{Rank} indicates the position of the default setting (lower is better); color encodes percentile: \textcolor{rankgreat}{\textbf{blue}} ($\geq$90\textsuperscript{th}), \textcolor{rankgood}{green} (61--89\textsuperscript{th}), \textcolor{rankmid}{orange} (26--60\textsuperscript{th}), \textcolor{rankbad}{\textbf{red}} ($\leq$25\textsuperscript{th}).}
\label{tab:default-position}
\setlength{\tabcolsep}{2.5pt}
\renewcommand{\arraystretch}{1.15}
\footnotesize
\begin{tabular}{@{}ll ccc cc c@{}}
\toprule
\textbf{Model} & \textbf{Metric} & \textbf{Default} & \textbf{Best} & \textbf{Worst} & $\boldsymbol{\Delta}$\textsubscript{\textbf{gap\_w}} & $\boldsymbol{\Delta}$\textsubscript{\textbf{gap\_d}} & \textbf{Rank} \\
\midrule
\multirow{4}{*}{\makecell[l]{GPT-OSS\\120B}}
 & VE pass@1  & .574 & .626 & .503 & .123 & .052 & \pctcell{46/109}{58} \\
 & VE pass@5  & .677 & .710 & .658 & .052 & .032 & \pctcell{47/109}{57} \\
 & RTL pass@1 & .383 & .575 & .319 & .255 & .192 & \pctcell{96/109}{12} \\
 & RTL pass@5 & .596 & .681 & .553 & .128 & .085 & \pctcell{51/109}{54} \\
\midrule
\multirow{4}{*}{\makecell[l]{Qwen-3.5\\397B}}
 & VE pass@1  & .619 & .845 & .600 & .245 & .226 & \pctcell{106/109}{3} \\
 & VE pass@5  & .710 & .942 & .710 & .232 & .232 & \pctcell{109/109}{0} \\
 & RTL pass@1 & .468 & .596 & .362 & .234 & .128 & \pctcell{61/109}{45} \\
 & RTL pass@5 & .575 & .681 & .575 & .106 & .106 & \pctcell{100/109}{9} \\
\midrule
\multirow{4}{*}{GLM-5}
 & VE pass@1  & .626 & .671 & .574 & .097 & .045 & \pctcell{61/109}{45} \\
 & VE pass@5  & .723 & .735 & .684 & .052 & .013 & \pctcell{4/109}{97} \\
 & RTL pass@1 & .489 & .617 & .447 & .170 & .128 & \pctcell{82/109}{26} \\
 & RTL pass@5 & .660 & .702 & .596 & .106 & .043 & \pctcell{28/109}{75} \\
\bottomrule
\end{tabular}
\end{table}

\subsubsection{Pass Rate–HQI Trade-off}
\begin{figure}[t]
  \centering
  \includegraphics[width=\linewidth]{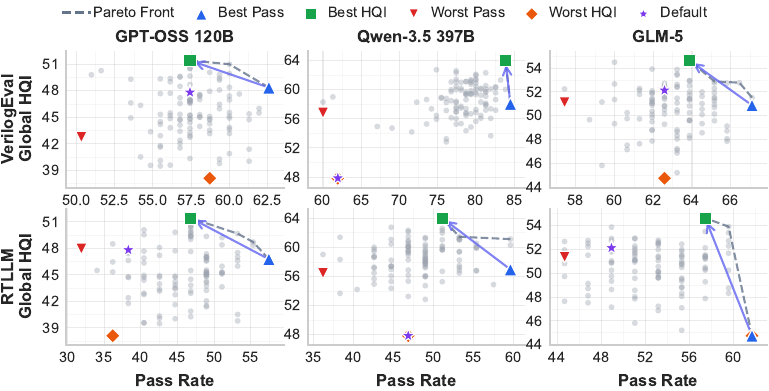}
  \caption{Scatter plots of Pass Rate vs. HQI across different LLMs and datasets. The Pareto frontier (dashed line) highlights the trade-off between functional correctness and comprehensive hardware quality.}
  \label{fig:passrate_hqi}
\end{figure}

Figure~\ref{fig:passrate_hqi} plots pass rate against Global HQI for all 109 configurations. The overall trend is positive: configurations that achieve higher pass rates generally yield higher HQI, confirming that functional correctness and synthesis quality are broadly aligned. However, the configuration that maximizes pass rate never exactly coincides with the one that maximizes HQI; the two optima occupy distinct positions along a Pareto frontier, and a subset of configurations deviate from the main trend, achieving strong pass rates with mediocre HQI or vice versa. This divergence is most visible for Qwen-3.5 397B on VerilogEval, where the best-pass and best-HQI settings sit at opposite ends of the trade-off curve. In practice, while pass rate remains a reasonable proxy for synthesis quality in most configurations, practitioners who require Pareto-optimal designs must treat hyperparameter selection as a multi-objective problem, calibrating not only for the model and benchmark but also for whether the deployment objective prioritizes functional correctness, post-synthesis quality, or a balance of both. The default configurations tend to cluster in the lower-left region of the scatter, far from the Pareto frontier, consistent with the ranking analysis in Table~\ref{tab:default-position} and reinforcing that out-of-the-box settings underperform on both objectives simultaneously.

\begin{figure*}[h]
    \centering
    \includegraphics[width=\textwidth]{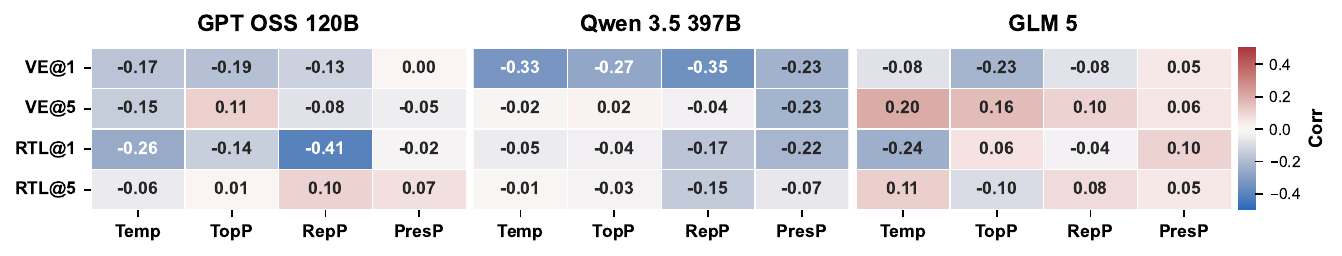}
    \caption{Hyperparameter--performance correlation across the three models.}
    \label{fig:hp_sensitivity}
\end{figure*}

\subsubsection{Hyperparameter Correlation}
Figure~\ref{fig:hp_sensitivity} maps the correlation between each decoding parameter and generation quality across the three models. No single parameter universally governs RTL output. Temperature and top\_p correlate negatively with pass@1 for GPT-OSS~120B and Qwen-3.5~397B, indicating that broader sampling introduces more noise than improvement under Verilog's strict syntactic constraints, whereas GLM-5 shows weaker and occasionally positive responses, suggesting greater tolerance to randomness. The penalty parameters reveal further architecture-specific patterns: Qwen-3.5~397B's strong negative correlation with presence penalty on VerilogEval suggests that forcing unseen tokens causes deviation from synthesizable logic, while GLM-5's mild positive response to repetition penalty may help suppress redundant logic blocks. GPT-OSS~120B exhibits intermediate behavior. These profiles caution against a ``best practice'' configuration; the next section reveals a fundamental limitation.

\subsubsection{Cross-Benchmark Non-Transferability}

The correlation patterns in Figure~\ref{fig:hp_sensitivity} often weaken or reverse between VerilogEval and RTLLM, raising the question of whether a configuration optimized on one benchmark retains its advantage on the other. To test this directly, all 108 configurations are ranked by pass@1 on each benchmark independently and Spearman's rank correlation $\rho$ between the two rankings. As shown in Table~\ref{tab:spearman}, the correlations are weak to negligible. 


GPT-OSS~120B shows a weak positive correlation that, while statistically significant, explains less than 6\% of the ranking variance. For Qwen-3.5~397B and GLM-5, the correlations are not statistically significant, meaning that a hyperparameter setting that ranks among the best on VerilogEval has essentially no predictive value for its ranking on RTLLM. This non-transferability is the strongest evidence against a universal ``recommended configuration'' for open-source LLMs in RTL generation. Practitioners must treat hyperparameter selection as a task-specific calibration, rather than a one-time decision.

\begin{table}[htbp]
\centering
\caption{Spearman's rank correlation ($\rho$) of hyperparameter configuration rankings between VerilogEval and RTLLM (pass@1, 108 configurations per model).}
\label{tab:spearman}
\begin{tabular}{@{} cc @{\hskip 2em} cc @{\hskip 2em} cc @{}}
\toprule
\multicolumn{2}{c}{\textbf{GPT-OSS 120B}} & \multicolumn{2}{c}{\textbf{Qwen-3.5 397B}} & \multicolumn{2}{c}{\textbf{GLM-5}} \\
\cmidrule(lr){1-2} \cmidrule(lr){3-4} \cmidrule(lr){5-6}
$\rho$ & $p$-value & $\rho$ & $p$-value & $\rho$ & $p$-value \\
\midrule
0.23 & 0.016 & 0.15 & 0.121 & $-$0.05 & 0.590 \\
\bottomrule
\end{tabular}
\end{table}

\subsubsection{Model Sensitivity on Hyperparameters}

\begin{figure}[htbp]
    \centering
    \includegraphics[width=0.8\linewidth]{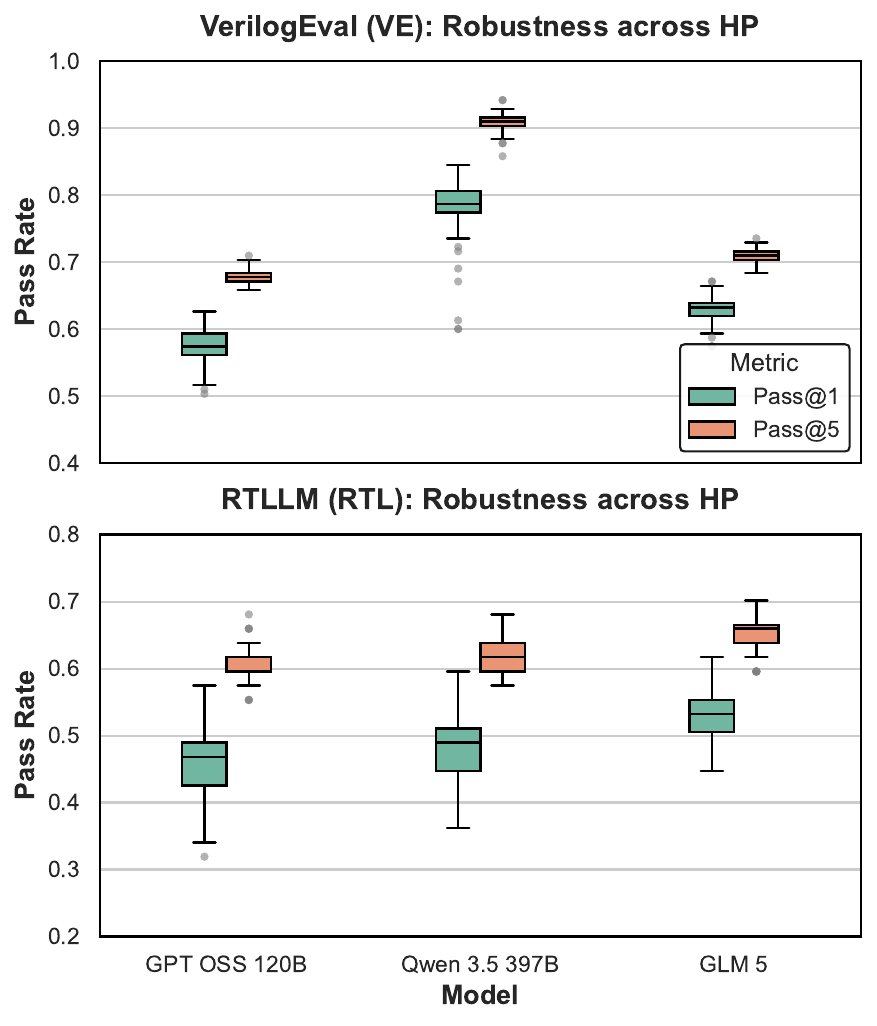}
    \caption{Distribution of pass rates under 108 hyperparameter configurations for each model.}
    \label{fig:hp_robustness}
    \vspace{-4mm}
\end{figure}

Figure~\ref{fig:hp_robustness} shows the distributional spread of pass rates under varying configurations. Pass@5 distributions remain compact across all three models, indicating stable underlying hardware knowledge, whereas pass@1 distributions exhibit wider spreads and downward outliers, confirming that first-attempt correctness is more sensitive to decoding choices. The effect is stronger on RTLLM, whose multi-module and protocol-level tasks~\cite{lu2024rtllm} amplify configuration impact: both GPT-OSS~120B and Qwen-3.5~397B show outliers approaching near-zero pass rates, demonstrating that a poorly configured top-tier model can underperform a well-configured smaller one.

GLM-5 maintains the tightest boxes on both benchmarks, safest when per-task tuning is infeasible. GPT-OSS~120B and Qwen-3.5~397B reach higher peaks but require careful configuration to avoid degradation. These distributions reinforce the central point: the same model can appear mediocre or competitive depending on its decoding configuration.


\section{Conclusion}
This work establishes that for open-source LLMs in RTL generation, \emph{how} a model is configured can matter more than \emph{which} model is chosen. Three findings support this conclusion at increasing levels of specificity. First, the 25.5\% pass-rate gap induced by hyperparameter variation within a single model exceeds the gap between many model families evaluated under default settings, and a well-tuned smaller model can surpass a poorly configured model with over 3$\times$ the parameters. Second, sensitivity to decoding configuration is architecture-specific: GLM-5 is robust across settings while GPT-OSS~120B and Qwen-3.5~397B exhibit double-digit swings, creating a robustness--ceiling trade-off. 
Third, optimal configurations are benchmark-specific: Spearman's rank correlation of configuration rankings across VerilogEval and RTLLM is near zero, ruling out ``recommended settings'' and requiring task-aware calibration.

These results have a direct methodological implication: benchmark comparisons of open-source LLMs conducted under default or unspecified hyperparameters conflate model quality with configuration quality. Reporting RTL generation results without specifying decoding parameters is incomplete, in the same way that reporting training results without specifying the learning rate would be. For practitioners deploying open-source LLMs in secure, on-premise EDA workflows, these findings indicate that a modest investment in per-task hyperparameter calibration can yield performance that rivals or exceeds much larger models, making open-source LLMs a viable solution when properly configured.

\bibliographystyle{IEEEtran}
\bibliography{refs}

@IEEEtranBSTCTL{IEEE:BSTcontrol,
  CTLuse_forced_etal       = "yes",
  CTLmax_names_forced_etal = "3",
  CTLnames_show_etal       = "2"
}

@misc{thakur2022benchmarkinglargelanguagemodels,
      title={Benchmarking Large Language Models for Automated Verilog RTL Code Generation}, 
      author={Shailja Thakur and Baleegh Ahmad and Zhenxing Fan and Hammond Pearce and Benjamin Tan and Ramesh Karri and Brendan Dolan-Gavitt and Siddharth Garg},
      year={2022},
      eprint={2212.11140},
      archivePrefix={arXiv},
      primaryClass={cs.PL},
      url={https://arxiv.org/abs/2212.11140}, 
}

@article{shao2024survey,
  title={Survey of different large language model architectures: Trends, benchmarks, and challenges},
  author={Shao, Minghao and Basit, Abdul and Karri, Ramesh and Shafique, Muhammad},
  journal={IEEE access},
  volume={12},
  pages={188664--188706},
  year={2024},
  publisher={IEEE}
}

@article{austin2021program,
  title={Program synthesis with large language models},
  author={Austin, Jacob and Odena, Augustus and Nye, Maxwell and Bosma, Maarten and Michalewski, Henryk and Dohan, David and Jiang, Ellen and Cai, Carrie and Terry, Michael and Le, Quoc and others},
  journal={arXiv preprint arXiv:2108.07732},
  year={2021}
}

@article{chen2021evaluating,
  title={Evaluating large language models trained on code},
  author={Chen, Mark and Tworek, Jerry and Jun, Heewoo and Yuan, Qiming and Pinto, Henrique Ponde De Oliveira and Kaplan, Jared and Edwards, Harri and Burda, Yuri and Joseph, Nicholas and Brockman, Greg and others},
  journal={arXiv preprint arXiv:2107.03374},
  year={2021}
}

@inproceedings{liu2023verilogeval,
  title={Verilogeval: Evaluating large language models for verilog code generation},
  author={Liu, Mingjie and Pinckney, Nathaniel and Khailany, Brucek and Ren, Haoxing},
  booktitle={2023 IEEE/ACM International Conference on Computer Aided Design (ICCAD)},
  pages={1--8},
  year={2023},
  organization={IEEE}
}

@inproceedings{wang2024llms,
  title={Llms and the future of chip design: Unveiling security risks and building trust},
  author={Wang, Zeng and Alrahis, Lilas and Mankali, Likhitha and Knechtel, Johann and Sinanoglu, Ozgur},
  booktitle={2024 IEEE Computer Society Annual Symposium on VLSI (ISVLSI)},
  pages={385--390},
  year={2024},
  organization={IEEE}
}

@inproceedings{zhu2024hot,
  title={Hot or cold? adaptive temperature sampling for code generation with large language models},
  author={Zhu, Yuqi and Li, Jia and Li, Ge and Zhao, YunFei and Jin, Zhi and Mei, Hong},
  booktitle={Proceedings of the AAAI Conference on Artificial Intelligence},
  volume={38},
  number={1},
  pages={437--445},
  year={2024}
}

@inproceedings{zhao2025mage,
  title={Mage: A multi-agent engine for automated rtl code generation},
  author={Zhao, Yujie and Zhang, Hejia and Huang, Hanxian and Yu, Zhongming and Zhao, Jishen},
  booktitle={2025 62nd ACM/IEEE Design Automation Conference (DAC)},
  pages={1--7},
  year={2025},
  organization={IEEE}
}

@inproceedings{wang2025verileaky,
  title={Verileaky: Navigating ip protection vs utility in fine-tuning for llm-driven verilog coding},
  author={Wang, Zeng and Shao, Minghao and Nabeel, Mohammed and Roy, Prithwish Basu and Mankali, Likhitha and Bhandari, Jitendra and Karri, Ramesh and Sinanoglu, Ozgur and Shafique, Muhammad and Knechtel, Johann},
  booktitle={2025 IEEE International Conference on LLM-Aided Design (ICLAD)},
  pages={100--107},
  year={2025},
  organization={IEEE}
}

@article{wang2025netdetox,
  title={NetDeTox: Adversarial and Efficient Evasion of Hardware-Security GNNs via RL-LLM Orchestration},
  author={Wang, Zeng and Shao, Minghao and Saha, Akashdeep and Karri, Ramesh and Knechtel, Johann and Shafique, Muhammad and Sinanoglu, Ozgur},
  journal={arXiv preprint arXiv:2512.00119},
  year={2025}
}

@misc{wolf2013yosys,
  title        = {Yosys Open {SYnthesis} Suite},
  author       = {Wolf, Clifford},
  howpublished = {\url{https://yosyshq.net/yosys/}},
  year         = {2013}
}

@misc{nangate45,
  title        = {The {NanGate} 45nm Open Cell Library},
  author       = {{Nangate Inc.}},
  howpublished = {\url{https://si2.org/}},
  year         = {2008}
}

@article{pinckney2025revisiting,
  title={Revisiting verilogeval: A year of improvements in large-language models for hardware code generation},
  author={Pinckney, Nathaniel and Batten, Christopher and Liu, Mingjie and Ren, Haoxing and Khailany, Brucek},
  journal={ACM Transactions on Design Automation of Electronic Systems},
  volume={30},
  number={6},
  pages={1--20},
  year={2025},
  publisher={ACM New York, NY}
}

@misc{fu2026synthesisintheloopevaluationllmsrtl,
      title={Synthesis-in-the-Loop Evaluation of LLMs for RTL Generation: Quality, Reliability, and Failure Modes}, 
      author={Weimin Fu and Zeng Wang and Minghao Shao and Ramesh Karri and Muhammad Shafique and Johann Knechtel and Ozgur Sinanoglu and Xiaolong Guo},
      year={2026},
      eprint={2603.11287},
      archivePrefix={arXiv},
      primaryClass={cs.AR},
      url={https://arxiv.org/abs/2603.11287}, 
}

@inproceedings{garcia2025turtle,
  title={Turtle: A unified evaluation of llms for rtl generation},
  author={Garcia-Gasulla, Dario and Kestor, Gokcen and Parisi, Emanuele and Albert{\'\i}-Binimelis, Miquel and Gutierrez, Cristian and Ghorab, Razine Moundir and Montenegro, Orlando and Homs, Bernat and Moreto, Miquel},
  booktitle={2025 ACM/IEEE 7th Symposium on Machine Learning for CAD (MLCAD)},
  pages={1--12},
  year={2025},
  organization={IEEE}
}

@article{liu2024rtlcoder,
  title={Rtlcoder: Fully open-source and efficient llm-assisted rtl code generation technique},
  author={Liu, Shang and Fang, Wenji and Lu, Yao and Wang, Jing and Zhang, Qijun and Zhang, Hongce and Xie, Zhiyao},
  journal={IEEE Transactions on Computer-Aided Design of Integrated Circuits and Systems},
  volume={44},
  number={4},
  pages={1448--1461},
  year={2024},
  publisher={IEEE}
}

@article{thakur2024verigen,
  title={Verigen: A large language model for verilog code generation},
  author={Thakur, Shailja and Ahmad, Baleegh and Pearce, Hammond and Tan, Benjamin and Dolan-Gavitt, Brendan and Karri, Ramesh and Garg, Siddharth},
  journal={ACM Transactions on Design Automation of Electronic Systems},
  volume={29},
  number={3},
  pages={1--31},
  year={2024},
  publisher={ACM New York, NY}
}

@article{lu2024openllmrtl,
  title={OpenLLM-RTL: Open Dataset and Benchmark for LLM-Aided Design RTL Generation},
  author={Lu, Yingjie and others},
  journal={arXiv preprint arXiv:2404.06117},
  year={2024}
}

@article{abdelatty2025pluto,
  title={Pluto: A Benchmark for Evaluating Efficiency of LLM-generated Hardware Code},
  author={Abdelatty, Manar and Nouh, Maryam and Rosenstein, Jacob K and Reda, Sherief},
  journal={arXiv preprint arXiv:2510.14756},
  year={2025}
}

@inproceedings{cui2024origen,
  title={Origen: Enhancing rtl code generation with code-to-code augmentation and self-reflection},
  author={Cui, Fan and Yin, Chenyang and Zhou, Kexing and Xiao, Youwei and Sun, Guangyu and Xu, Qiang and Guo, Qipeng and Liang, Yun and Zhang, Xingcheng and Song, Demin and others},
  booktitle={Proceedings of the 43rd IEEE/ACM International Conference on Computer-Aided Design},
  pages={1--9},
  year={2024}
}

@inproceedings{lu2024rtllm,
  title={Rtllm: An open-source benchmark for design rtl generation with large language model},
  author={Lu, Yao and Liu, Shang and Zhang, Qijun and Xie, Zhiyao},
  booktitle={2024 29th Asia and South Pacific Design Automation Conference (ASP-DAC)},
  pages={722--727},
  year={2024},
  organization={IEEE}
}

\end{document}